\documentclass[letter,bibyear]{aa} 

\usepackage{natbib}
\usepackage{graphicx}
\usepackage{txfonts}

\newcommand{\jms}{J.~Mol.~Spectrosc.}   
\newcommand{\jpc}{J.~Phys.~Chem.}  
\newcommand{\jmst}{J.~Mol.~Struct.}

\newcommand{\kms}{km s$^{-1}$}
\newcommand{\diez}{10$^{10}$\,cm$^{-2}$}
\newcommand{\once}{10$^{11}$\,cm$^{-2}$}
\newcommand{\doce}{10$^{12}$\,cm$^{-2}$}

\begin{document}

\title{Discovery of  HCCCO and C$_5$O in TMC-1 with the QUIJOTE$^1$ line survey\thanks{Based on observations carried out
with the Yebes 40m telescope (projects 19A003,
20A014, 20D023, and 21A011) and the Institut de Radioastronomie Millim\'etrique (IRAM) 30m telescope. The 40m
radiotelescope at Yebes Observatory is operated by the Spanish Geographic Institute
(IGN, Ministerio de Transportes, Movilidad y Agenda Urbana). IRAM is supported by INSU/CNRS
(France), MPG (Germany), and IGN (Spain).}}

\author{
J.~Cernicharo\inst{1},
M.~Ag\'undez\inst{1},
C.~Cabezas\inst{1},
B.~Tercero\inst{2,3},
N.~Marcelino\inst{3},
R.~Fuentetaja\inst{1}
JR.~Pardo\inst{1},
P.~deVicente\inst{2}
}

\institute{Grupo de Astrof\'isica Molecular, Instituto de F\'isica Fundamental (IFF-CSIC),
C/ Serrano 121, 28006 Madrid, Spain. \email jose.cernicharo@csic.es
\and Centro de Desarrollos Tecnol\'ogicos, Observatorio de Yebes (IGN), 19141 Yebes, Guadalajara, Spain.
\and Observatorio Astron\'omico Nacional (OAN, IGN), Madrid, Spain.
}

\date{Received; accepted}

\abstract{We report on the detection, for the first time in space, of the radical HCCCO and of
pentacarbon monoxide, C$_5$O. The derived column densities are (1.6$\pm$0.2)$\times$\once and
(1.5$\pm$0.2)$\times$\diez, respectively. We have also analysed the data for all the molecular
species of the families HC$_n$O and C$_n$O within our QUIJOTE$^1$'s line survey.
Upper limits are obtained for HC$_4$O, HC$_6$O, C$_4$O,
and C$_6$O. We report a robust detection of HC$_5$O and HC$_7$O based on 14 and
12 rotational lines detected with a signal-to-noise ratio $\ge$30 and $\ge$5, respectively.
The derived N(HC$_3$O)/N(HC$_5$O) abundance ratio is 0.09$\pm$0.03, while N(C$_3$O)/N(C$_5$O) is
80$\pm$2, and N(HC$_5$O)/N(HC$_7$O) is 2.2$\pm$0.3. 
As opposed to the cyanopolyyne family, HC$_{2n+1}$N, which shows a continuous decrease in the
abundances with increasing $n$, the C$_n$O and HC$_n$O species show a clear abundance maximum for $n$=3 and 5,
respectively. 
They also show an odd and even abundance alternation, with odd values of n being the most abundant, 
which is reminiscent of the behaviour of C$_n$H radicals, where in that case species with even values 
of $n$ are more abundant.
We explored the formation of these species through two mechanisms 
previously proposed, which are based on radiative associations between C$_n$H$_m^+$ ions 
with CO and reactions of C$_n^-$ and C$_n$H$^-$ anions with O atoms, and we found that several 
species, such as C$_5$O, HC$_4$O, and HC$_6$O, are significantly overestimated. Our understanding 
of how these species are formed is  incomplete as of yet. Other routes based on neutral-neutral reactions 
such as those of C$_n$ and C$_n$H carbon chains with O, OH, or HCO, could be behind the formation 
of these species.}


\titlerunning{HC$_n$O and C$_n$O species in TMC-1}
\authorrunning{Cernicharo et al.}

\maketitle

\section{Introduction}
Recent observations of the molecular cloud TMC-1 
with the Yebes 40m \citep{Cernicharo2021a} and Green Bank \citep{McGuire2018}
radio telescopes have shown that this source presents a paramount challenge to 
our understanding of the chemical processes in cold prestellar cores.
The detection of cyclopentadiene and indene \citep{Cernicharo2021b}, ortho-benzyne \citep{Cernicharo2021a},
the cyano
derivatives of cyclopentadiene, benzene and naphthalene \citep{McGuire2018,McCarthy2021,
Lee2021,McGuire2021}, the ethynyl derivatives of cyclopentandiene \citep{Cernicharo2021c},
together with the discovery of extremely abundant hydrocarbons such as 
propargyl \citep{Agundez2021a}, vinyl acetylene \citep{Cernicharo2021d}, and allenyl acetylene \citep{Cernicharo2021e}
suggest that key chemical processes involving hydrocarbons (radical, neutral, 
and/or ionic species) were not considered in previous
chemical networks. Most of these species were not predicted to have a significant abundance.
Hence, a new chemistry is emerging from these observations of TMC-1.

TMC-1 is also a source that is particularly rich in sulphur bearing molecules. In addition
to the well-known species CS, CCS, and C$_3$S in this source \citep{Saito1987,Yamamoto1987}, 
eight new S-bearing species have been discovered in the last year
with the QUIJOTE line survey: HCCCS$^+$, NCS, HCCS, H$_2$CCS, H$_2$CCCS, C$_4$S, HCSCN, and HCSCCH
\citep{Cernicharo2021f,Cernicharo2021g,Cernicharo2021h}. In addition, C$_5$S, which has previously only been detected in evolved stars, has also been detected in TMC-1 \citep{Cernicharo2021g}. Also complex O-bearing molecules, 
such as CH$_2$CHCHO, CH$_2$CHOH, HCOOCH$_3$, and
CH$_3$OCH$_3$, have been detected \citep{Agundez2021b} in this cloud. Moreover, the
protonated species of tricarbon monoxide and ketene, HC$_3$O$^+$ and CH$_3$CO$^+$, have also been detected 
 recently \citep{Cernicharo2020a,Cernicharo2021i}. On the other hand, one of the most intriguing 
results concerning O-bearing
species has been the detection of the long carbon chain radicals HC$_5$O and HC$_7$O which are particularly
abundant 
\citep{McGuire2017,Cordiner2017}. Therefore, similar molecules of the type C$_n$O and HC$_n$O 
could be present in TMC-1. The detection of these species could provide insights on the oxygen
chemistry in this source.

In this Letter, we report the discovery of the radical \mbox{HCCCO} (trans-propynonyl) and of the O-bearing carbon chain 
C$_5$O (pentacarbon monoxide). Line surveys with the Yebes 40m telescope (QUIJOTE\footnote{\textbf{Q}-band \textbf{U}ltrasensitive \textbf{I}nspection \textbf{J}ourney 
to the \textbf{O}bscure \textbf{T}MC-1 \textbf{E}nvironment}; see \citealt{Cernicharo2021a}) 
and the IRAM 30m radiotelescope have been used for this work. 
We report a detailed study
of the HC$_n$O and C$_n$O families of O-bearing species. The formation of these species is 
investigated with the aid of a chemical model. A robust detection
of HC$_5$O and HC$_7$O is also presented.

\begin{figure}[]
\centering
\includegraphics[scale=0.65,angle=0]{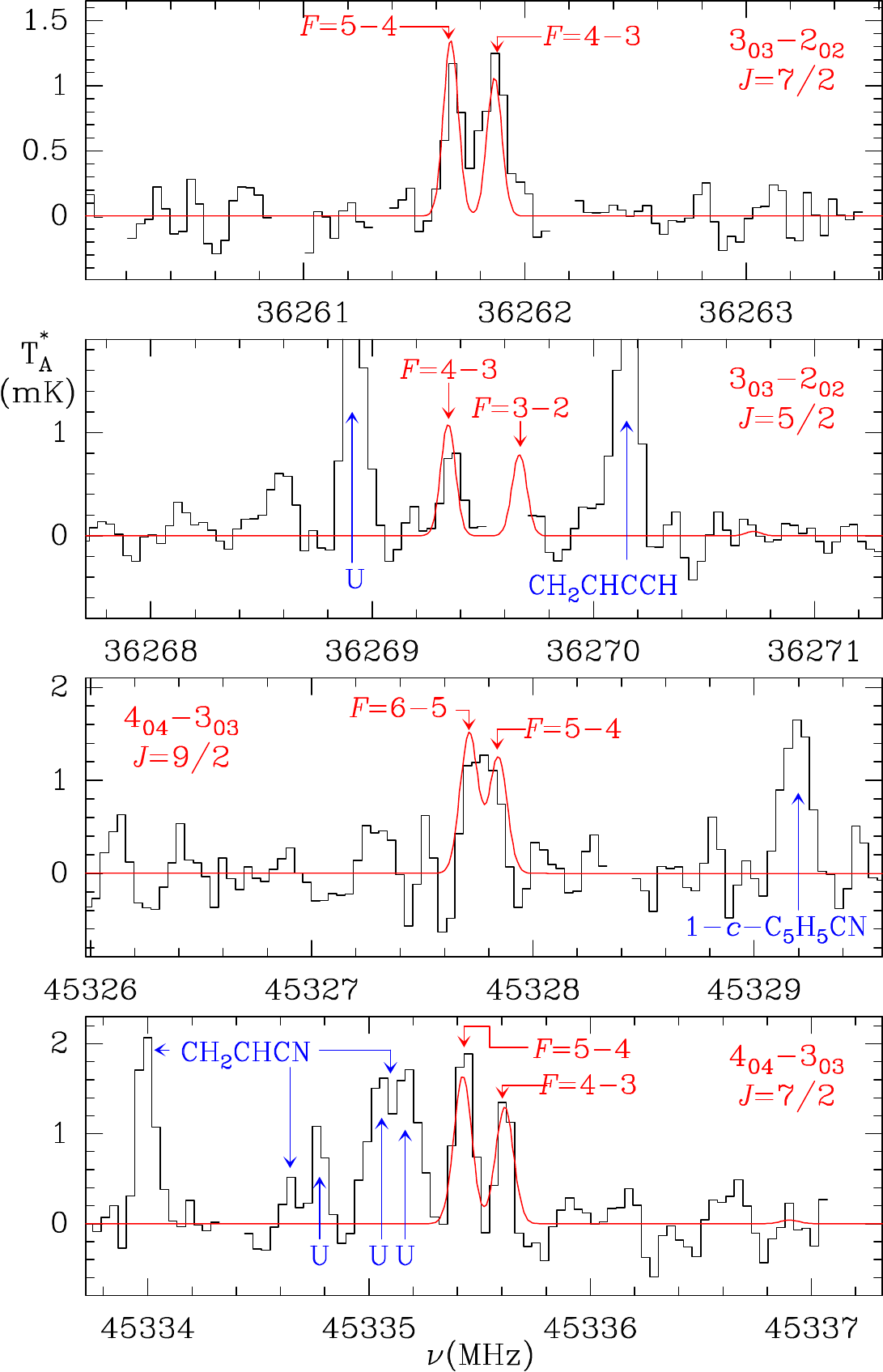}
\caption{Observed transitions of HCCCO in TMC-1.
The abscissa corresponds to the rest frequency of the lines. Frequencies and intensities for the observed lines
are given in Table \ref{obs_line_parameters}.
The ordinate is the antenna temperature, corrected for atmospheric and telescope losses, in milli Kelvin.
The quantum numbers of each transition are indicated
in the corresponding panel. The red line shows the computed synthetic spectrum for this species for T$_r$=7 K and
a column density of 1.3$\times$\once. Blank channels correspond to negative features 
produced in the folding of the frequency switching data.
}
\label{fig_hc3o}
\end{figure}

\section{Observations}
\label{observations}
New receivers, built within the Nanocosmos project\footnote{\texttt{https://nanocosmos.iff.csic.es/}}
and installed at the Yebes 40m radiotelescope, were used
for the observations of TMC-1
($\alpha_{J2000}=4^{\rm h} 41^{\rm  m} 41.9^{\rm s}$ and $\delta_{J2000}=
+25^\circ 41' 27.0''$). A detailed description of the system is 
given by \citet{Tercero2021}.
The receiver consists of two cold high electron mobility transistor amplifiers covering the
31.0-50.3 GHz band with horizontal and vertical             
polarisations. Receiver temperatures in the runs achieved during 2020 vary from 22 K at 32 GHz
to 42 K at 50 GHz. Some power adaptation in the down-conversion chains have reduced
the receiver temperatures during 2021 to 16\,K at 32 GHz and 30\,K at 50 GHz.
The backends are $2\times8\times2.5$ GHz fast Fourier transform spectrometers
with a spectral resolution of 38.15 kHz,
providing the whole coverage of the Q-band in both polarisations. 
All observations were performed in the frequency 
switching mode with frequency throws of 8 and 10 MHz. 
A detailed description of the QUIJOTE line survey is
provided in \citet{Cernicharo2021a}.
The main beam efficiency varies from 0.6 at
32 GHz to 0.43 at 50 GHz. 
Pointing corrections were derived from nearby quasars and SiO masers,
and errors were always within 2-3$''$. The telescope 
beam size is 56$''$ and 31$''$ at 31 and 50 GHz, respectively.
The intensity scale used in this work, that is to say the antenna temperature
($T_A^*$), was calibrated using two absorbers at different temperatures and the
atmospheric transmission model (ATM; \citealt{Cernicharo1985, Pardo2001}).
Calibration uncertainties were adopted to be 10~\%.

The IRAM 30m data come from a line survey performed towards TMC-1 and B1 and the observations
have been described by \citet{Marcelino2007} and \cite{Cernicharo2012b}.
All data were analysed using the GILDAS package\footnote{\texttt{http://www.iram.fr/IRAMFR/GILDAS}}.

\section{Results}
\label{results}
The sensitivity of our observations towards TMC-1, between 0.19-0.35 mK (1$\sigma$), is much larger
than
previously published line surveys of this source at the same frequencies \citep{Kaifu2004}. 
In fact, it has been possible to detect many individual lines from molecular species
that were previously reported by only stacking techniques \citep{Marcelino2021,Cernicharo2021b,Cernicharo2021c}. 
Line identification in this work has been performed using the MADEX catalogue \citep{Cernicharo2012},
the Cologne Database of Molecular Spectroscopy (CDMS) catalogue (\citealt{Muller2005}), and the
Jet Propulsion Laboratory (JPL) catalogue (JPL; \citealt{Pickett1998}).

\subsection{First detection of trans-propynonyl (HCCCO)}
\label{hccco}
Rotational spectroscopy for this radical is available from the microwave \citep{Chen1996}
and millimetre and sub-millimetre domains \citep{Cooksy1992a,Cooksy1992b}. Frequency predictions 
as provided by the CDMS  catalogue \citep{Muller2005} are rather accurate in the 31-50 GHz domain. They
have been implemented in the MADEX code. The dipole moment adopted for this species is
$\mu_a$=2.39\,D \citep{Cooksy1995}. 
The ground electronic state of the molecule corresponds to a $^2A'$ quasilinear species.
Due to the large $A$ rotational constant, $\sim$261 GHz \citep{Cooksy1992a}, all rotational
levels with energies below 12.2 K have $K_a$=0 up to $J$=12. The first $K_a$=1 level is the 1$_{11}$ $J$=1/2 $F$=1,
which has an energy of 12.4 K. Hence, the strongest lines for a cold dark cloud with T$_K$=10 K
are those connecting levels with $K_a$=0, with weaker emission from $K_a$=1 transitions. 
Two of the $K_a$=0 rotational transitions, the 3$_{03}$-2$_{02}$ and 4$_{04}$-3$_{03}$, are in the domain
of the QUIJOTE line survey and are predicted to have intensities above 1 mK for
column densities $\ge$10$^{11}$ cm$^{-2}$.
These transitions exhibit a fine ($J=N\pm1/2$) and hyperfine structure, and they have been detected with QUIJOTE.
Fig. \ref{fig_hc3o} shows the observed lines and the derived line parameters provided 
in Table \ref{obs_line_parameters}. The averaged v$_{LSR}$ of the lines is {5.75$\pm$0.04\,\kms, which 
agrees within 2$\sigma$ with the value obtained for the cyanopolyynes family of 5.83$\pm$0.03\,\kms\, \citep{Cernicharo2020b}.

The two observed rotational transitions have very close upper level energies and thus cannot 
provide a reasonable estimate
for T$_{rot}$. Moreover, no collisional rates are available for this species. Hence,
in order to derive a column density, we need to assume a rotational temperature.
We have assumed the value derived for C$_3$O in Appendix \ref{Ap_CCCO} (T$_{rot}$=7\,K). Adopting a source of uniform
brightness and a diameter of 80$''$ \citep{Fosse2001}, which means that the source practically 
fills the beam at all of the frequencies of the survey,
we derived a column density for HCCCO of (1.3$\pm$0.2)$\times$\once. The column densities of other
members of the HC$_n$O family are derived in Appendix \ref{Ap_HCnO} and are given in Table \ref{col_densities}.
We note that HC$_5$O appears to be the most abundant species of the HC$_n$O family of O-bearing carbon chains, with a relative abundance
of 1.3$\pm$0.3, 1.8$\pm$0.3, 11$\pm$3, and 2.2$\pm$0.3 with respect to
HCO, HCCO, HC$_3$O, and HC$_7$O, respectively. Also, HC$_4$O and HC$_6$O have been searched, but only upper limits
have been obtained (see Appendix \ref{Ap_HC5O_HC7O}).

\begin{figure}[]
\centering
\includegraphics[scale=0.65,angle=0]{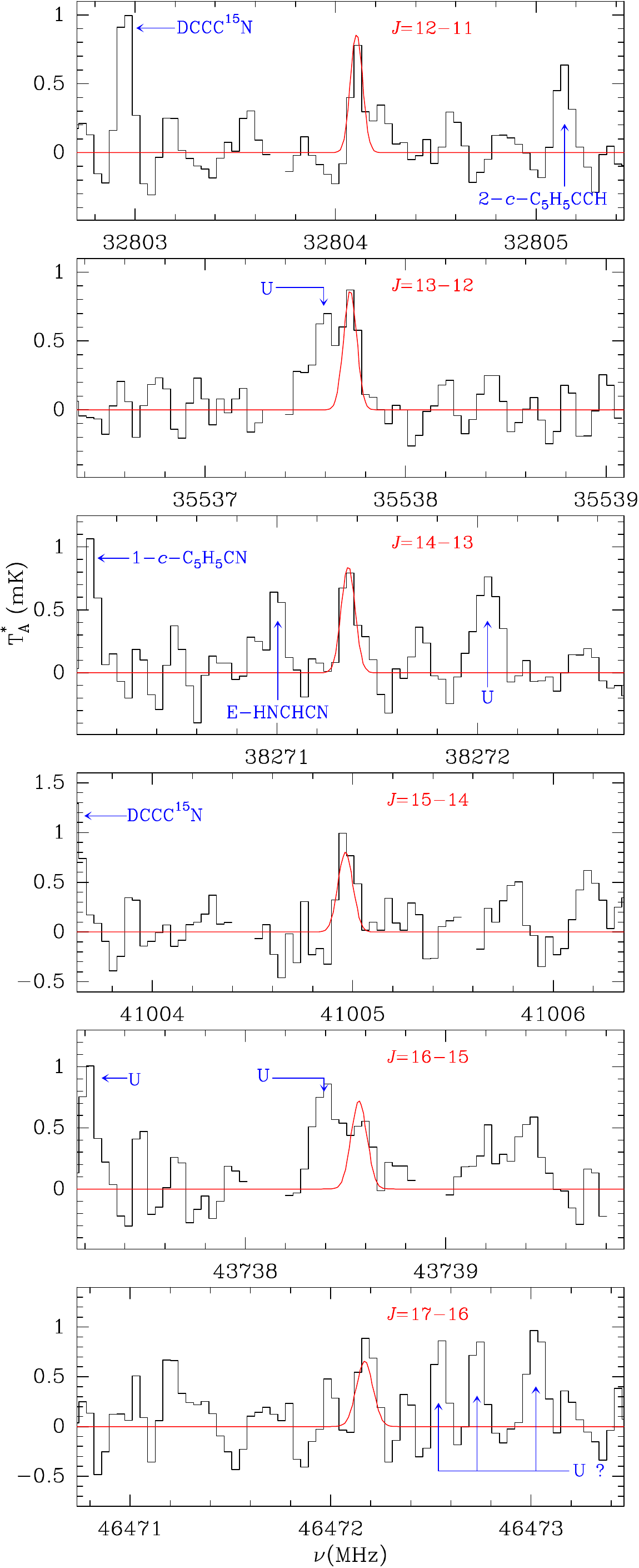}
\caption{Similar to Fig. \ref{fig_hc3o}, but for the observed transitions of C$_5$O in TMC-1.
The red line shows the computed synthetic spectrum for this species for T$_r$=10 K and
a column density of 1.5$\times$\diez. 
}
\label{fig_c5o}
\end{figure}

\subsection{First detection of pentacarbon monoxide (C$_5$O)}
\label{detection_c5o}
The O-bearing chains CCO and C$_3$O have been observed in TMC-1 with large
column densities (see Appendices \ref{Ap_CCO} and \ref{Ap_CCCO}). 
Hence, other similar chains such as C$_4$O, C$_5$O, and C$_6$O could be present in this cloud.

We note that C$_5$O has been observed in the microwave laboratory up to 24.6 GHz and $J_{max}$=9
by \citet{Ogata1995}. The uncertainty of the frequency measurements is 1 kHz. 
We adopted a dipole moment of 4.06\,D as derived by \citet{Botschwina1995} from
ab initio calculations.
Six lines of C$_5$O from $J=12-11$ up to $J=17-16$ are clearly detected at a 
v$_{LSR}$ of 5.38-5.66 \kms. They are
shown in Fig. \ref{fig_c5o}. All the lines of C$_5$O are
shifted by 15-40 kHz with respect the predictions if a v$_{LSR}$ of 5.75\,\kms\, is
adopted. This velocity is the average value of the v$_{LSR}$ velocities 
observed for the lines of HCCCO, CCO, C$_3$O, HC$_5$O, and
HC$_7$O (see Table \ref{obs_line_parameters} and Appendices \ref{Ap_HCnO}, \ref{Ap_CnO}). 
Consequently, we adopted a velocity of 5.75 km\,s$^{-1}$ to derive the rest frequencies of
the observed lines of C$_5$O. We merged them with the laboratory data of \citet{Ogata1995} 
to derive improved rotational and distortion constants. The obtained values are $B$=1366.847016$\pm$0.000058 MHz and
$D$=0.03413$\pm$0.00045 kHz, while the laboratory data alone provide
$B$=1366.847112$\pm$0.000065 MHz and $D$=0.03509$\pm$0.00056 kHz. 

A rotational diagram of the observed lines of C$_5$O provides a rotational temperature of 
10$\pm$0.5 K and a column density of (1.5$\pm$0.2)$\times$10$^{10}$ cm$^{-2}$. In order to check the excitation
conditions for the rotational lines of this species, we assumed that the collisional rates
of C$_5$O could be similar to those of HC$_5$N with $p$-H$_2$
(kindly provided by Francois Lique, private communication). For a linewidth of 0.6 km\,s$^{-1}$, a volume
density of 4$\times$10$^4$ cm$^{-3}$ \citep{Cernicharo1987,Fosse2001}, 
and a kinetic temperature of 10 K, large velocity gradient (LVG) 
calculations performed with the MADEX code indicate that all rotational lines up to $J_u$=20
will have excitation temperatures close to thermalisation (between 9.5 and 10 K). This result
is similar to that obtained for HC$_5$N and its isotopologues from the data in our line survey
\citep{Cernicharo2020b}. The synthetic spectra computed with a rotational temperature of 10 K and the column
density derived from the rotational diagram
are shown in Fig. \ref{fig_c5o} and are in excellent agreement with observations.

\begin{figure*}
\centering
\includegraphics[width=0.93\textwidth]{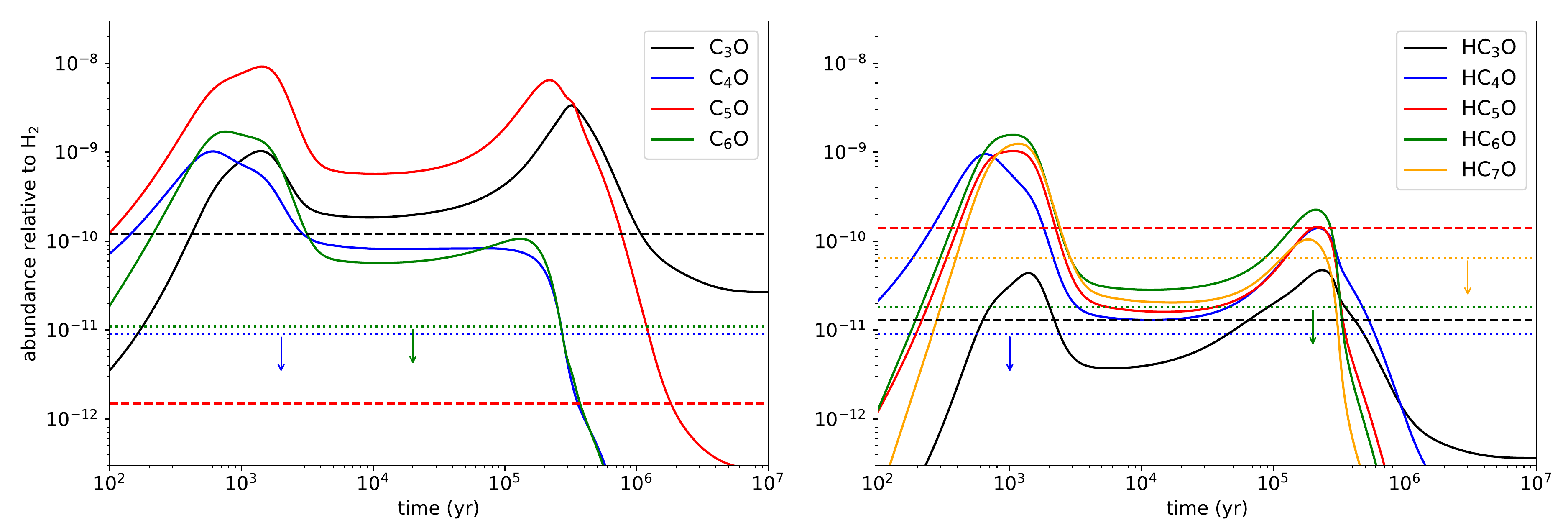}
\caption{Calculated fractional abundances for several C$_n$O and HC$_n$O species as a function of time. Observed abundances in \mbox{TMC-1} are indicated by horizontal dashed lines, while horizontal dotted lines represent upper limits for non-detected species.}
\label{fig:abun}
\end{figure*}

Assuming a column density for H$_2$ of 10$^{22}$ cm$^{-2}$ \citep{Cernicharo1987}, 
we derived an abundance for C$_5$O of 2$\times$10$^{-12}$. This is 50$\pm$15 and 80$\pm$20 times lower than that
of CCO and C$_3$O, respectively (see Appendices \ref{Ap_CCO}, \ref{Ap_CCCO}, and Table \ref{col_densities}). 
We searched for the species C$_4$O and C$_6$O (see Appendix \ref{Ap_C4O_C6O}),
but only upper limits have been obtained. They are given in Table \ref{col_densities}.
It is important to note that C$_3$O appears to be the most abundant species of the C$_n$O family (n$\ge$2).

\section{Discussion} 
\label{discussion}

The derived abundance for the different members of the C$_n$O ($n\ge2$) and HC$_n$O
($n\ge1$) families permit one to study the possible routes for their formation. It is worth noting
that HC$_5$O is the more abundant species of both series, and that the heavy species
HC$_7$O is only slightly lower in abundance. This is a peculiar result indeed as it is
clearly opposite to the behaviour of the HC$_{2n+1}$N family, where the abundance decreases
with $n$, and the different molecules are formed by the consecutive addition of CCH or CN 
\citep{Cernicharo1987b,Agundez2008}. 
Nevertheless,\ the peculiarities of HC$_5$O and HC$_7$O are not restricted to the physical conditions
of interstellar clouds. In the laboratory experiments devoted to the rotational characterisation
of these molecules, \citet{Mohamed2005} reported that HC$_5$O was
the most abundant species generated in their discharges for different precursors. 
In the Stardust and AROMA experimental setups of the Nanocosmos 
project \citep{Lidia2020,Santoro2020,Sabbah2017}, it has been observed that in experiments 
using pure carbon as a seed for the
growth of nanoparticles, masses associated with HC$_5$O and HC$_7$O 
are clearly detected, with that of HC$_7$O being rather prominent and comparable
in intensity to C$_7$H$_2$ \citep{Lidia2020}. In similar experiments, but adding C$_2$H$_2$
to the growth chamber, HC$_3$O, HC$_5$O, and HC$_7$O are also detected.
The source of the oxygen for the formation of these molecules is probably related
to air contamination (O$_2$) during transportation of the samples \citep{Lidia2020,Santoro2020}. No other
oxidised molecules were detected in these experiments. Although the physical conditions between
the interstellar clouds and these experiments are very different, it seems that these molecules 
are easily formed under a variety of conditions. 

Two main mechanisms have been proposed to account for the formation of C$_n$O and HC$_n$O species in interstellar space. The first involves the radiative association of C$_{n-1}$H$^+$, C$_{n-1}$H$_2^+$, and C$_{n-1}$H$_3^+$ ions with CO \citep{Adams1989}, followed by dissociative recombination with electrons,
\begin{subequations} \label{reac:cation}
\begin{align}
&\rm C_{\textit n-1}H^+ + CO \rightarrow HC_{\textit n}O^+ + {\textit h\nu}, \label{reac:cation_a} \\
&\rm C_{\textit n-1}H_2^+ + CO \rightarrow H_2C_{\textit n}O^+ + {\textit h\nu}, \nonumber \\
&\rm C_{\textit n-1}H_3^+ + CO \rightarrow H_3C_{\textit n}O + {\textit h\nu}, \nonumber \\
&\rm HC_{\textit n}O^+ + e^- \rightarrow C_{\textit n}O + H, \label{reac:cation_b} \\
&\rm H_2C_{\textit n}O^+ + e^- \rightarrow HC_{\textit n}O + H, \nonumber \\
&\rm H_3C_{\textit n}O^+ + e^- \rightarrow HC_{\textit n}O + H + H. \nonumber
\end{align}
\end{subequations}
The second consists of direct formation through reactions of C$_n^-$ and C$_n$H$^-$ anions with O 
atoms \citep{Eichel2007,Cordiner2012},
\begin{flalign} \label{reac:anion}
&\rm C_{\textit n}^- + O \rightarrow C_{\textit n}O + e^-, \\
&\rm C_{\textit n}H^- + O \rightarrow HC_{\textit n}O + e^-. \nonumber
\end{flalign}
The two mechanisms have problems reproducing some of the abundances observed 
for C$_n$O and HC$_n$O in \mbox{TMC-1}. \cite{McGuire2017} implemented the mechanism 
(\ref{reac:cation}) in a chemical model and found that the abundance HC$_4$O is overestimated 
by at least two orders of magnitude, while including the two mechanisms (\ref{reac:cation}) 
and (\ref{reac:anion}) resulted in an overestimation of the abundances of HC$_6$O, C$_6$O, 
and C$_7$O by at least one or two orders of magnitude \citep{Cordiner2012,Cordiner2017}.

Here we included the reactions relevant for mechanisms (\ref{reac:cation}) and (\ref{reac:anion}) 
in a chemical model of a cold dark cloud similar to those presented in previous studies of 
\mbox{TMC-1} (e.g. \citealt{Agundez2021a}). We adopted rate coefficients for radiative associations 
in scheme (\ref{reac:cation_a}) from \cite{Adams1989}; for dissociative recombinations in scheme 
(\ref{reac:cation_b}) from the estimations of \cite{Loison2017} for HC$_3$O$^+$, H$_2$C$_3$O$^+$, 
and H$_3$C$_3$O$^+$; and for reactions of negative ions with O atoms in scheme (\ref{reac:anion}) 
from \cite{Eichel2007} and \cite{Cordiner2012}. We assume that C$_n$O and HC$_n$O species react 
with the most abundant cations (H$^+$, C$^+$, He$^+$, HCO$^+$, and H$_3$O$^+$) and with C atoms, 
but we consider that only open shell species (all HC$_n$O species and C$_n$O with $n$ even) react 
with H, N, and O atoms.

The results are shown in Fig.~\ref{fig:abun}. The observed abundances of HC$_3$O, HC$_5$O, 
and HC$_7$O are reasonably reproduced. However, HC$_4$O and HC$_6$O are 
overestimated by one order of magnitude. That is, the chemical model does not make a strong 
differentiation between HC$_n$O species with
an even $n$ and an odd $n$, in contrast to observations 
which indicate that HC$_n$O species with $n$ odd are favoured. Concerning the carbon chains C$n$O, 
the chemical model predicts a clear differentiation between C$_n$O with an even $n$ and an odd $n$, as 
a result of the different reactivity with H, N, and O atoms. However, the calculated abundances 
are too large. This is especially dramatic in the case of C$_5$O, whose calculated abundance is 
almost four orders of magnitude too high. The chemical mechanisms (\ref{reac:cation}) and 
(\ref{reac:anion}) are too efficient at producing O-bearing carbon chains, and they lack specificity 
between an even $n$ and and an odd $n$ for HC$_n$O species. The assumptions behind these mechanisms must 
be revised, in particular the rate coefficients adopted for the radiative associations in 
scheme (\ref{reac:cation_a}) and the branching ratios leading to C$_n$O in the reactions between 
C$_n^-$ and O atoms in scheme (\ref{reac:anion}). In addition, it would be desirable to know the 
true low-temperature reactivity of these O-bearing carbon chains with neutral atoms.

The overabundance of some of the C$_n$O and HC$_n$O species in the chemical model including the 
chemical mechanisms (\ref{reac:cation}) and (\ref{reac:anion}), together with the fact that HC$_3$O, 
HC$_5$O, and HC$_7$O are observed in the Nanocosmos experiments, suggest that other chemical 
routes may be behind the formation of these species. In particular, the reactions of C$_n$ and 
C$_n$H carbon chains with O atoms and the radicals OH and HCO are potential sources of C$_n$O 
and HC$_n$O species. The alternation in the abundances of consecutive C$_n$H radicals, 
with those with $n$ even being more abundant that those with $n$ odd, could produce, in 
their reaction with HCO, a similar behaviour for the HC$_n$O species in which case those 
with $n$ odd would be more abundant than those with $n$ even. However, little is known about these reactions.

\section{Conclusions}

We detected, for the first time in space, the species HC$_3$O and C$_5$O. In addition, we have provided an 
exhaustive list of abundances and upper limits for O-bearing carbon chains C$_n$O and HC$_n$O in 
\mbox{TMC-1}. The species with $n$ odd are found to be more abundant than those with $n$ even. 
In particular, C$_3$O and HC$_5$O are especially abundant. Formation mechanisms for these species 
based on radiative associations of C$_n$H$_m^+$ ions with CO and reactions between C$_n^-$ and 
C$_n$H$^-$ anions with O atoms reproduce some features, but result in abundances that are  too
large for 
some species such as C$_5$O, HC$_4$O, and HC$_6$O. Alternative chemical routes for these species 
involving neutral-neutral reactions should be explored.

\begin{table}
\caption{Column densities for relevant O-bearing species in TMC-1}
\label{col_densities}
\centering
\tiny
\begin{tabular}{{|l|c|c|c|}}
\hline
Molecule        & $N^a$                      &$T_{rot}$$^b$& $X^c$                      \\ 
\hline                                                                                         
C$_2$O          &  (7.5$\pm$0.3)$\times$10$^{11}$ &  7-10.0$^d$ & 7.5$\times$10$^{-11}$       \\ 
C$_3$O          &  (1.2$\pm$0.2)$\times$10$^{12}$ & 7.0$\pm$1.0 & 1.2$\times$10$^{-10}$       \\ 
C$_4$O          & $\le$9.0$\times$10$^{10}$       &  10.0       &$\le$9.0$\times$10$^{-12}$   \\ 
C$_5$O          & (1.5$\pm$0.2)$\times$10$^{10}$  &10.0$\pm$0.5 & 1.5$\times$10$^{-12}$       \\ 
C$_6$O          & $\le$1.1$\times$10$^{11}$       &  10.0       &$\le$1.1$\times$10$^{-11}$   \\ 
HCO             &  (1.1$\pm$0.1)$\times$10$^{12}$&   7.0       &1.1$\times$10$^{-10}$         \\ 
HC$_2$O         &  (7.7$\pm$0.7)$\times$10$^{11}$ &  7.0        &7.7$\times$10$^{-11}$        \\ 
HC$_3$O         &  (1.3$\pm$0.2)$\times$10$^{11}$ &  7.0        &1.3$\times$10$^{-11}$        \\ 
HC$_4$O         & $\le$9.0$\times$10$^{10}$       &  10.0       &$\le$9.0$\times$10$^{-12}$   \\ 
HC$_5$O         &  (1.4$\pm$0.2)$\times$10$^{12}$ &  10.0       & 1.4$\times$10$^{-10}$       \\ 
HC$_6$O         & $\le$1.8$\times$10$^{11}$       &  10.0       &$\le$1.8$\times$10$^{-11}$   \\ 
HC$_7$O         & (6.5$\pm$0.5)$\times$10$^{11}$   &  10.0       & 6.5$\times$10$^{-11}$      \\ 
\hline
\end{tabular}
\tablefoot{\\
        \tablefoottext{a}{Column density in cm$^{-2}$. Upper limits correspond to 3$\sigma$ values.}\\
        \tablefoottext{b}{Rotational temperature. When the uncertainty is not provided, it corresponds
     to an assumed value.}\\
        \tablefoottext{c}{Relative abundance to H$_2$, assuming a total column density of
    molecular hydrogen of 10$^{22}$ cm$^{-2}$ \citep{Cernicharo1987}.}\\
        \tablefoottext{d}{Column density derived from a large velocity gradient model (see Appendix
\ref{Ap_CCO}).}\\
}
\end{table}
\normalsize

\begin{acknowledgements}

We thank Ministerio de Ciencia e Innovaci\'on of Spain (MICIU) for funding support through projects
PID2019-106110GB-I00, PID2019-107115GB-C21 / AEI / 10.13039/501100011033, and
PID2019-106235GB-I00. We also thank ERC for funding
through grant ERC-2013-Syg-610256-NANOCOSMOS. M.A. thanks MICIU for grant RyC-2014-16277.
We would like to thank C. Joblin for her usefull comments on the
laboratory experiments of formation of carbon dust analogues in which HC$_3$O, HC$_5$O and HC$_7$O 
are observed. We would like also to thank Evelyne Roueff and Octavio Roncero for useful comments
on the possible reactions leading to the formation of HC$_n$O.
\end{acknowledgements}

\normalsize

\onecolumn
\begin{appendix}
\section{Observed line parameters}
\label{line_parameters}
Line parameters for the different molecules studied in this work were obtained by fitting a Gaussian line
profile to the observed data. A window of $\pm$ 15 \kms\, around the v$_{LSR}$ of the source was
considered for each transition. The derived line parameters for all the molecular species 
studied in this work are given in Table \ref{obs_line_parameters}.


\small
\begin{longtable}{cccllrc}
\caption{Observed line parameters for the species studied in  this work} \label{obs_line_parameters}\\
\hline \hline
Transition         &$\nu_{rest}$~$^a$    & $\int T_A^* dv$~$^b$ & \multicolumn{1}{c}{$\Delta v$~$^c$} & \multicolumn{1}{c}{$\Delta$v$^d$}    & \multicolumn{1}{c}{$T_A^*$$^e$} &  \\
                   &  (MHz)             & (mK\,km\,s$^{-1}$)   & \multicolumn{1}{c}{(km\,s$^{-1}$)}  & \multicolumn{1}{c}{(km\,s$^{-1}$)}    & \multicolumn{1}{c}{(mK)}       &  \\
\hline
\endfirsthead
\caption{continued.}\\
\hline \hline
Transition        &$\nu_{rest}$~$^a$    & $\int T_A^* dv$~$^b$ & \multicolumn{1}{c}{$\Delta v$~$^c$} & \multicolumn{1}{c}{$\Delta$v$^d$}    & \multicolumn{1}{c}{$T_A^*$$^e$} &  \\
                   &  (MHz)             & (mK\,km\,s$^{-1}$)   & \multicolumn{1}{c}{(km\,s$^{-1}$)}  & \multicolumn{1}{c}{(km\,s$^{-1}$)}    & \multicolumn{1}{c}{(mK)}       &  \\
\hline
\endhead
\hline
\endfoot
\hline
\multicolumn{7}{c}{\bf HCCCO$^1$} \\
$4_{0,4}-3_{0,3}$  9/2-7/2 5-4&36261.666$\pm$0.004& 0.90$\pm$0.12& 5.73$\pm$0.04& 0.74$\pm$0.11& 1.14$\pm$0.15& \\
$4_{0,4}-3_{0,3}$  9/2-7/2 4-3&36261.863$\pm$0.004& 1.41$\pm$0.14& 5.83$\pm$0.05& 1.15$\pm$0.15& 1.14$\pm$0.15& \\
$4_{0,4}-3_{0,3}$  7/2-5/2 4-3&36269.344$\pm$0.003& 0.75$\pm$0.14& 5.66$\pm$0.08& 0.82$\pm$0.16& 0.85$\pm$0.15& \\
$4_{0,4}-3_{0,3}$  7/2-5/2 3-2&36269.666$\pm$0.003&              &              &              &              & A\\  
$5_{0,5}-4_{0,4}$ 11/2-9/2 6-5&45327.712$\pm$0.004& 0.63$\pm$0.19& 5.74$\pm$0.20& 0.60         & 0.99$\pm$0.22& B\\  
$5_{0,5}-4_{0,4}$ 11/2-9/2 5-4&45327.843$\pm$0.004& 0.61$\pm$0.18& 5.92$\pm$0.20& 0.60         & 0.96$\pm$0.22& B\\  
$5_{0,5}-4_{0,4}$  9/2-7/2 5-4&45335.424$\pm$0.004& 1.35$\pm$0.15& 5.79$\pm$0.04& 0.62$\pm$0.07& 2.06$\pm$0.22& \\
$5_{0,5}-4_{0,4}$  9/2-7/2 4-3&45335.613$\pm$0.004& 0.73$\pm$0.13& 5.82$\pm$0.05& 0.44$\pm$0.09& 1.57$\pm$0.22& \\
\hline
\multicolumn{7}{c}{\bf CCO$^2$} \\
$1_1-0_1$          & 32623.449$\pm$0.007& 2.71$\pm$0.30& 5.53$\pm$0.05& 0.87$\pm$0.11&  2.94$\pm$0.27& \\
$2_1-1_1$          & 32738.613$\pm$0.005& 1.23$\pm$0.23& 5.73$\pm$0.06& 0.70$\pm$0.13&  1.65$\pm$0.24& \\
$2_3-1_2$          & 45826.734$\pm$0.002& 9.11$\pm$0.47& 5.75$\pm$0.07& 0.75$\pm$0.03& 11.36$\pm$0.46& \\
$2_2-1_1$          & 46182.187$\pm$0.002& 2.73$\pm$0.30& 5.70$\pm$0.03& 0.49$\pm$0.07&  5.26$\pm$0.44& \\
$4_5-3_4$          & 92227.870$\pm$0.003& 6.34$\pm$0.35& 5.71$\pm$0.03& 0.53$\pm$0.09& 11.34$\pm$1.53& \\
$4_4-3_3$          & 92363.257$\pm$0.004& 3.37$\pm$0.47& 5.64$\pm$0.05& 0.32$\pm$0.05& 11.06$\pm$1.37& \\
$4_3-3_2$          & 92718.775$\pm$0.003& 4.31$\pm$0.38& 5.73$\pm$0.03& 0.64$\pm$0.07&  6.29$\pm$0.80& \\
\hline
\multicolumn{7}{c}{\bf C$_3$O$^3$} \\
4- 3              & 38486.891$\pm$0.001&38.76$\pm$0.34& 5.74$\pm$0.01& 0.59$\pm$0.01& 61.78$\pm$0.43& \\
 5- 4              & 48108.474$\pm$0.001&40.83$\pm$0.15& 5.74$\pm$0.01& 0.65$\pm$0.01& 58.61$\pm$0.23& \\
 8- 7              & 76972.587$\pm$0.001&30.31$\pm$0.53& 5.75$\pm$0.01& 0.55$\pm$0.01& 51.56$\pm$1.10& \\
 9- 8              & 86593.685$\pm$0.001&17.30$\pm$1.40& 5.75$\pm$0.02& 0.41$\pm$0.04& 39.26$\pm$3.00& \\
10- 9              & 96214.614$\pm$0.001& 8.47$\pm$0.22& 5.75$\pm$0.01& 0.50$\pm$0.02& 16.07$\pm$0.55& \\
11-10              &105835.358$\pm$0.002& 4.85$\pm$0.71& 5.78$\pm$0.03& 0.38$\pm$0.07& 12.04$\pm$2.30& \\
\hline
\multicolumn{7}{c}{\bf C$_5$O$^{3,4}$} \\
12-11              & 32804.100$\pm$0.010& 0.69$\pm$0.21& 5.75         & 0.57$\pm$0.19&  0.68$\pm$0.17& \\ 
13-12              & 35527.730$\pm$0.010& 0.53$\pm$0.15& 5.75         & 0.51$\pm$0.20&  0.97$\pm$0.18& \\ 
14-13              & 38271.346$\pm$0.010& 0.55$\pm$0.13& 5.75         & 0.67$\pm$0.19&  0.78$\pm$0.18& \\ 
15-14              & 41004.959$\pm$0.020& 0.79$\pm$0.14& 5.75         & 0.69$\pm$0.13&  1.07$\pm$0.22& \\ 
16-15              & 43738.562$\pm$0.030& 0.57$\pm$0.19& 5.75         & 0.95$\pm$0.33&  0.56$\pm$0.23& \\ 
17-16              & 46472.169$\pm$0.030& 0.42$\pm$0.12& 5.75         & 0.56$\pm$0.11&  0.96$\pm$0.26& \\ 
\hline
\multicolumn{7}{c}{\bf HCO$^{1}$} \\
$1_{01}-0_{00}$ 3/2-1/2 2-1&86670.760$\pm$0.060&42.52$\pm$1.97& 5.94$\pm$0.02& 0.66$\pm$0.04& 60.69$\pm$3.91& \\
$1_{01}-0_{00}$ 3/2-1/2 1-0&86708.360$\pm$0.040&              &              &              &               &C\\
$1_{01}-0_{00}$ 1/2-1/2 1-1&86777.460$\pm$0.040&23.94$\pm$1.67& 5.90$\pm$0.02& 0.60$\pm$0.06& 37.57$\pm$3.10& \\
$1_{01}-0_{00}$ 1/2-1/2 0-1&86805.780$\pm$0.100&11.84$\pm$1.70& 5.95$\pm$0.04& 0.58$\pm$0.10& 19.25$\pm$3.27& \\
\hline
\multicolumn{7}{c}{\bf HCCO$^1$} \\
$2_{02}-1_{01}$ 5/2-3/2 3-2&43317.667$\pm$0.004 &5.80$\pm$0.19& 5.71$\pm$0.02   &0.68$\pm$0.03& 8.00$\pm$0.22& \\
$2_{02}-1_{01}$ 5/2-3/2 2-1&43321.145$\pm$0.004 &4.22$\pm$0.27& 5.66$\pm$0.04   &0.92$\pm$0.08& 4.30$\pm$0.22& \\
$2_{02}-1_{01}$ 3/2-1/2 2-1&43329.542$\pm$0.003 &5.07$\pm$0.26& 5.69$\pm$0.02   &0.73$\pm$0.04& 6.56$\pm$0.22& \\
$2_{02}-1_{01}$ 3/2-1/2 1-0&43335.463$\pm$0.004 &2.23$\pm$0.25& 5.70$\pm$0.08   &0.98$\pm$0.13& 2.14$\pm$0.22& \\
$2_{02}-1_{01}$ 5/2-3/2 2-2&43336.861$\pm$0.004 &0.98$\pm$0.25& 5.66$\pm$0.08   &0.63$\pm$0.15& 1.45$\pm$0.22& \\
$2_{02}-1_{01}$ 3/2-3/2 1-1&43337.304$\pm$0.006 &1.54$\pm$0.23& 5.67$\pm$0.08   &0.75$\pm$0.15& 1.93$\pm$0.22& \\
$4_{04}-3_{03}$ 9/2-7/2 5-4&86642.342$\pm$0.006& 3.16$\pm$1.00& 5.64$\pm$0.08   &0.22$\pm$0.11&13.71$\pm$3.80& \\
$4_{04}-3_{03}$ 9/2-7/2 4-3&86643.848$\pm$0.005& 4.27$\pm$1.09& 5.67$\pm$0.08   &0.26$\pm$0.11&15.31$\pm$3.80& \\
$4_{04}-3_{03}$ 7/2-5/2 4-3&86655.831$\pm$0.005& 7.23$\pm$1.45& 5.69$\pm$0.08   &0.40$\pm$0.13&16.97$\pm$3.80& \\
$4_{04}-3_{03}$ 7/2-5/2 3-2&86657.485$\pm$0.005& 2.96$\pm$0.97& 5.64$\pm$0.12   &0.24$\pm$0.12&11.80$\pm$3.80& \\ 
\hline
\multicolumn{7}{c}{\bf HC$_5$O$^5$} \\
25/2-23/2 $e$ 13-12        &32267.964$\pm$0.002& 3.86$\pm$0.42& 5.75$\pm$0.03   &0.67$\pm$0.06& 5.38$\pm$0.19& \\
25/2-23/2 $e$ 12-11        &32268.049$\pm$0.002& 4.08$\pm$0.42& 5.75$\pm$0.04   &0.75$\pm$0.06& 5.10$\pm$0.19& \\
25/2-23/2 $f$ 13-12        &32271.760$\pm$0.002& 4.11$\pm$0.29& 5.67$\pm$0.03   &0.75$\pm$0.05& 5.14$\pm$0.19& \\
25/2-23/2 $f$ 12-11        &32271.848$\pm$0.002& 3.21$\pm$0.30& 5.72$\pm$0.02   &0.65$\pm$0.06& 5.38$\pm$0.19& \\
27/2-25/2 $e$ 14-13        &34849.461$\pm$0.002& 4.15$\pm$0.11& 5.74$\pm$0.01   &0.60         & 6.49$\pm$0.19&D\\
27/2-25/2 $e$ 13-12        &34849.540$\pm$0.002& 3.18$\pm$0.11& 5.73$\pm$0.02   &0.60         & 4.99$\pm$0.19&D\\
27/2-25/2 $f$ 14-13        &34853.387$\pm$0.002& 3.55$\pm$0.09& 5.71$\pm$0.01   &0.60         & 5.55$\pm$0.19&D\\
27/2-25/2 $f$ 13-12        &34854.469$\pm$0.002& 2.79$\pm$0.10& 5.67$\pm$0.01   &0.60         & 4.37$\pm$0.19&D\\
29/2-27/2 $e$ 15-14        &37430.945$\pm$0.003& 3.67$\pm$0.15& 5.74$\pm$0.02   &0.60         & 5.74$\pm$0.21&D\\
29/2-27/2 $e$ 14-13        &37431.020$\pm$0.003& 3.42$\pm$0.14& 5.73$\pm$0.02   &0.60         & 5.35$\pm$0.21&D\\
29/2-27/2 $f$ 15-14        &37435.011$\pm$0.003& 3.70$\pm$0.14& 5.72$\pm$0.02   &0.60         & 5.79$\pm$0.21&D\\
29/2-27/2 $f$ 14-13        &37435.088$\pm$0.003& 3.29$\pm$0.14& 5.70$\pm$0.02   &0.60         & 5.14$\pm$0.21&D\\
31/2-29/2 $e$              &40012.451$\pm$0.004& 6.25$\pm$0.19& 5.73$\pm$0.01   &0.94$\pm$0.03& 6.26$\pm$0.25& \\
31/2-29/2 $f$              &40016.669$\pm$0.004& 6.13$\pm$0.19& 5.73$\pm$0.02   &0.95$\pm$0.03& 6.05$\pm$0.25& \\
33/2-31/2 $e$              &42593.906$\pm$0.005& 4.76$\pm$0.17& 5.75$\pm$0.02   &0.89$\pm$0.04& 5.03$\pm$0.23& \\
33/2-31/2 $f$              &42598.284$\pm$0.005& 5.42$\pm$0.17& 5.75$\pm$0.02   &0.90$\pm$0.03& 5.69$\pm$0.23& \\
35/2-33/2 $e$              &45175.347$\pm$0.005& 4.88$\pm$0.23& 5.79$\pm$0.02   &0.87$\pm$0.04& 5.31$\pm$0.28& \\
35/2-33/2 $f$              &45179.895$\pm$0.005& 5.51$\pm$0.26& 5.77$\pm$0.02   &1.00$\pm$0.06& 5.20$\pm$0.28& \\
37/2-35/2 $e$              &47746.772$\pm$0.007& 3.59$\pm$0.33& 5.82$\pm$0.04   &0.71$\pm$0.09& 4.36$\pm$0.30& \\
37/2-35/2 $f$              &47761.502$\pm$0.007& 3.89$\pm$0.36& 5.72$\pm$0.04   &0.72$\pm$0.07& 5.05$\pm$0.30& \\
\hline
\multicolumn{7}{c}{\bf HC$_7$O$^5$} \\
57/2-55/2 $e$              &31274.822$\pm$0.003& 0.90$\pm$0.25& 5.81$\pm$0.08   &0.84$\pm$0.21& 1.01$\pm$0.19& \\
57/2-55/2 $f$              &31276.500$\pm$0.003& 1.05$\pm$0.15& 5.89$\pm$0.06   &0.78$\pm$0.12& 1.28$\pm$0.19& \\
59/2-57/2 $e$              &32372.184$\pm$0.003&              &                 &             &              &E\\     
59/2-57/2 $f$              &32373.871$\pm$0.003& 1.29$\pm$0.10& 5.89$\pm$0.04   &0.87$\pm$0.08& 1.39$\pm$0.19& \\
61/2-59/2 $e$              &33469.542$\pm$0.004& 1.02$\pm$0.11& 6.00$\pm$0.04   &0.74$\pm$0.02& 1.30$\pm$0.20& \\
61/2-59/2 $f$              &33471.241$\pm$0.004&              &                 &             &              &E\\     
63/2-61/2 $e$              &34566.898$\pm$0.004&              &                 &             &              &F\\     
63/2-61/2 $f$              &34568.607$\pm$0.004& 0.53$\pm$0.08& 5.73$\pm$0.08   &0.65$\pm$0.16& 0.76$\pm$0.20& \\
65/2-63/2 $e$              &35664.244$\pm$0.005& 0.37$\pm$0.07& 5.78$\pm$0.18   &0.58$\pm$0.21& 0.59$\pm$0.18& \\
65/2-63/2 $f$              &35665.972$\pm$0.005& 0.54$\pm$0.07& 5.68$\pm$0.25   &1.22$\pm$0.32& 0.43$\pm$0.18& \\
67/2-65/2 $e$              &36761.602$\pm$0.005& 0.43$\pm$0.12& 5.49$\pm$0.16   &1.11$\pm$0.35& 0.41$\pm$0.21& \\
67/2-65/2 $f$              &36763.333$\pm$0.005& 0.59$\pm$0.10& 5.69$\pm$0.07   &0.79$\pm$0.12& 0.70$\pm$0.21& \\
69/2-67/2 $e$              &37858.949$\pm$0.005& 0.33$\pm$0.10& 6.00$\pm$0.12   &0.43$\pm$0.25& 0.71$\pm$0.23& \\
69/2-67/2 $f$              &37860.691$\pm$0.005& 0.55$\pm$0.13& 6.04$\pm$0.09   &0.74$\pm$0.19& 0.70$\pm$0.23& \\
\hline
\end{longtable}
\tablefoot{\\
\tablefoottext{a}{Rest frequency of the transition. See text for the laboratory data used
for the frequency predictions of each molecular species.}\\
\tablefoottext{b}{Integrated line intensity in mK\,km\,s$^{-1}$.}\\
\tablefoottext{c}{v$_{LSR}$ in \kms.}\\

\tablefoottext{d}{Linewidth at half intensity derived by fitting a Gaussian function to
the observed line profile (in km\,s$^{-1}$).}\\
\tablefoottext{e}{Antenna temperature in milli Kelvin.}\\
\tablefoottext{1}{Quantum numbers are $N, K_a, K_c, J$, and $F$.}\\
\tablefoottext{2}{Quantum numbers are $N$ and $J$.}\\
\tablefoottext{3}{The quantum number is $J$.}\\
\tablefoottext{4}{Improved rest frequencies were derived assuming a v$_{LSR}$ of 5.75 \kms\,
(see Sect. \ref{detection_c5o}).}\\
\tablefoottext{5}{Quantum numbers are $J$, $\Lambda$-doubling component, and $F$. 
If the hyperfine structure is not resolved in our data, then $F$ is not
given. In this case, the adopted rest frequency corresponds to the 
average of those of the hyperfine components.}\\
\tablefoottext{A}{This line is contaminated by the frequency switching negative feature of the transition 
$4_{0,4}-3_{0,3}$ 9/2-7/2 5-4 of the same species after data folding since they are 
separated by exactly 8.0 MHz, i.e. the frequency throw used in the observations.}\\
\tablefoottext{B}{This doublet appears unresolved in our data (see Fig. \ref{fig_hc3o}). The linewidth
has been fixed to 0.6 \kms. The resulting frequencies and intensities have slightly larger uncertainties than
those of other transitions for this molecule.}\\
\tablefoottext{C}{The $J$=3/2-1/2 $F$=1-0 component is heavily blended with the $J$=15-14
transition of C$_3$S avoiding a reliable determination of the line parameters (see Fig. \ref{fig_hco}).}\\
\tablefoottext{D}{The hyperfine components are only partially resolved. To obtain reliable parameters
for each hyperfine component, their linewidths are fixed to 0.6\,\kms.}\\
\tablefoottext{E}{Heavily affected by a negative feature produced in the 
folding of the frequency switching data (see Fig. \ref{fig_hc7o}).}\\
\tablefoottext{F}{This transition is heavily blended with the $J$=31-30
transition of HC$_6$$^{13}$CN at 34566.972 MHz, which has an intensity of 2.1 mK (see Fig. \ref{fig_hc7o})}\\

}
\normalsize
\twocolumn

\section{HC$_n$O species}
\label{Ap_HCnO}
\subsection{HCO}
\label{Ap_HCO}
The HCO radical has an electronic ground state $^2A'$
and has been observed in the laboratory by different authors
\citep{Saito1972,Austin1974,Pickett1978,Blake1984}. 
Frequency predictions
are available in the JPL catalogue \citep{Pickett1998}. We
implemented the molecule in the MADEX catalogue and searched for its lines
in our 3mm line survey of TMC-1. No lines of HCO appear in the QUIJOTE line
survey. Four hyperfine components of the 1$_{01}$-0$_{00}$ transition are detected
and they are shown in Fig. \ref{fig_hco}. Their line parameters are given
in Table \ref{obs_line_parameters}. One of these lines is heavily blended 
with the $J$=15-14 line of C$_3$S. The observed velocities are systematically
shifted by 0.15 \kms\, with respect the velocity of CCO, C$_3$O, HCCO, and HC$_3$O.
This is a result of the low accuracy, 40-100 kHz, of the laboratory measurements for this
transition (see Table \ref{obs_line_parameters} and \citet{Pickett1978}).
We adopted a rotational temperature
of 7 K and derived a column density of (1.1$\pm$0.1)$\times$\doce.

\begin{figure*}[]
\centering
\includegraphics[scale=0.60,angle=0]{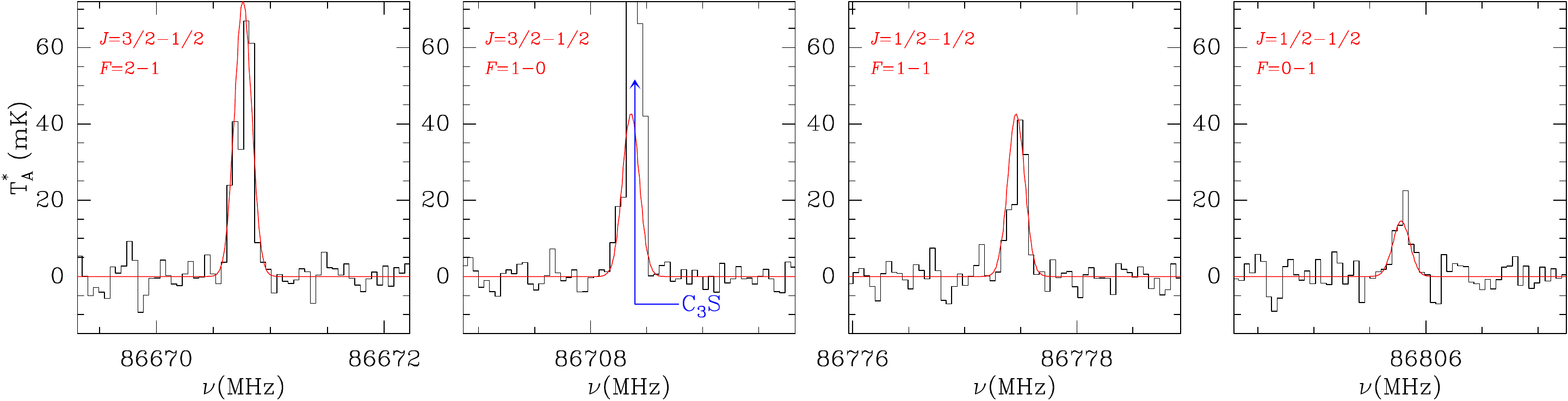}
\caption{Observed lines of the $1_{01}-0_{00}$ transition of HCO in TMC-1.
The abscissa corresponds to the rest frequency of the lines. The
v$_{LSR}$, linewidth, and integrated intensity of each line are provided in Table \ref{obs_line_parameters}.
The ordinate is the antenna temperature, corrected for atmospheric and telescope losses, in 
milli Kelvin. The noise of this spectrum is 3.7 mK.
The quantum numbers for each transition are indicated. 
The red line shows the computed synthetic spectrum for this species (see Appendix \ref{Ap_HCO}). 
}
\label{fig_hco}
\end{figure*}
\subsection{HCCO}
\label{Ap_HCCO}
We note that HCCO has an electronic ground state $^2A''$ and its
rotational transitions with $K_a$=0 have been observed in the laboratory by 
different authors \citep{Endo1987,Ohshima1993,Chantzos2019}. The dipole moment of
the molecule has been estimated through ab initio calculations to be $\mu_a$=1.59 D,
with a small component along the $b$ axis of the molecule \citep{Szalay1996,Muller2005}.
Predictions for the transitions of this molecule are provided in the CDMS catalogue
\citep{Muller2005} and have been implemented in MADEX. The molecule has been detected towards
several cold dark clouds by \citet{Agundez2015}.
The $N=2-1$ transition has been observed with the QUIJOTE line survey. The six fine and hyperfine
components are shown in Fig. \ref{fig_hcco}. The derived line parameters are given in 
Table \ref{obs_line_parameters}.
Two transitions of HCCO, the $N=4-3$ and $N=5-4$, have also been observed in our 3mm data.
The four strongest hyperfine components of the $N=4-3$ are detected at the 3$\sigma$ level
and their paramaters are given in Table \ref{obs_line_parameters}. For the $N=5-4$ transition,
only upper limits of 11 mK (3$\sigma$) are obtained. Adopting the rotational temperature of
CCO (7 K), and a source of uniform brightness with a diameter of 80$''$
\citep{Fosse2001}, we derived a column density of  (7.7$\pm$0.7)$\times$\once. 
The synthetic spectrum for the $N=2-1$ transition is shown in red in Fig. \ref{fig_hcco} and
matches the observed lines rather well. The predicted intensities
for the $N=4-3$ lines are in agreement with the derived parameters for the four
strongest hyperfine components of this transition.

\begin{figure*}[]
\centering
\includegraphics[scale=0.75,angle=0]{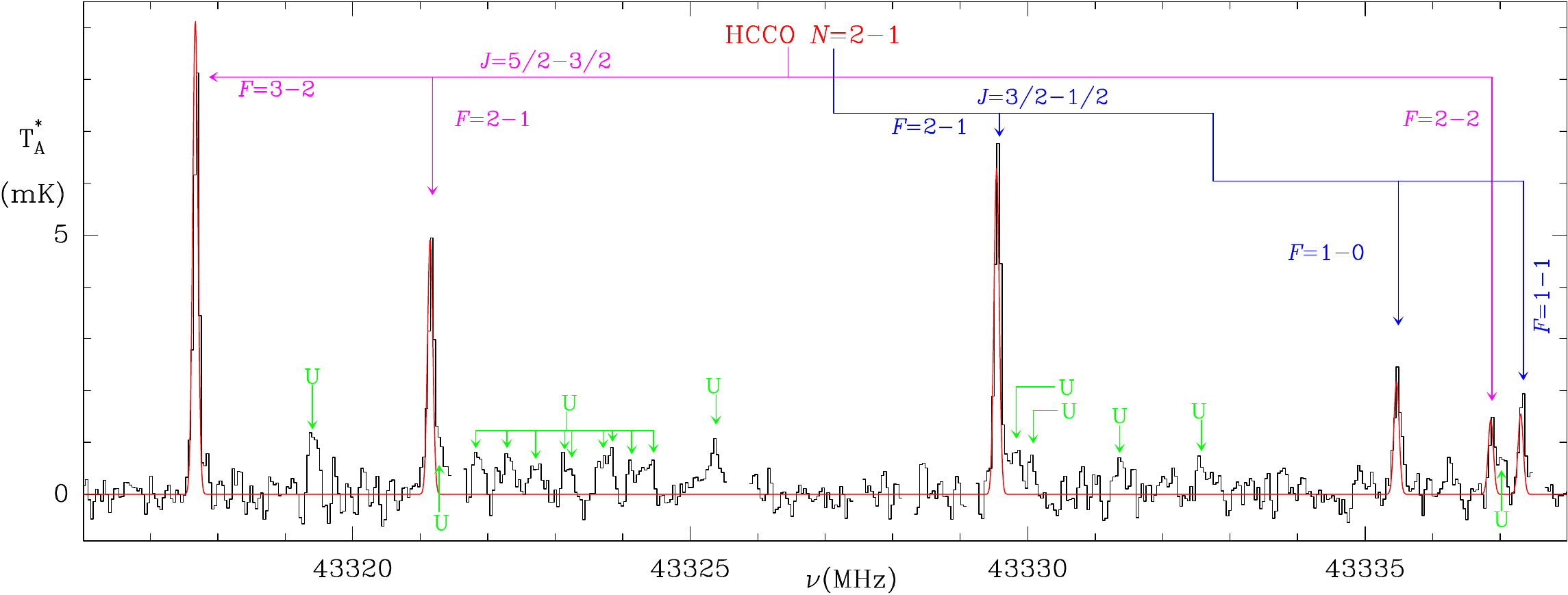}
\caption{Observed lines of the $2_{02}-1_{01}$ transition of HCCO in TMC-1.
The abscissa corresponds to the rest frequency of the lines. The
v$_{LSR}$, linewidth, and integrated intensity of each line are given in Table \ref{obs_line_parameters}.
The ordinate is the antenna temperature, corrected for atmospheric and telescope losses, in 
milli Kelvin. The noise of this spectrum is 0.22 mK.
The quantum numbers for each transition are indicated. The hyperfine components of
the fine structure transitions
$J$=5/2-3/2 and $J$=3/2-1/2 are indicated in violet and blue, respectively.
The red line shows the computed synthetic spectrum for this species (see Appendix \ref{Ap_HCCO}).
Blank channels correspond to negative features produced in the folding of the frequency
switching data. 
}
\label{fig_hcco}
\end{figure*}

\subsection{HC$_5$O and HC$_7$O}
\label{Ap_HC5O_HC7O}
It is important to note that HC$_5$O is a linear molecule with a $^2\Pi_r$ state and its $^2\Pi_{1/2}$
rotational spectrum has been observed in the laboratory by
\citet{Mohamed2005}, who also provided an estimate of the dipole moment of 2.16\,D from ab initio calculations.
The molecule has been detected in TMC-1 by \citet{McGuire2017} through the detection of 
the two $\Lambda$-components, each one splitted in two hyperfine components,
of the $J$=17/2-15/2 rotational transition at 21.85 GHz. In Fig. \ref{fig_hc5o} we report the
two $\Lambda$-components of seven rotational transitions from $J$=25/2-23/2 up to $J$=37/2-35/2.
The signal-to-noise ratio for each individual line is larger than 30. Line parameters for the
observed lines are given in Table \ref{obs_line_parameters}. The derived velocities of the observed
transitions are $\sim$5.75\,\kms, that is similar to those of the lines of C$_3$O.
We adopted a rotational temperature
of 10 K and derived a column density for HC$_5$O of (1.4$\pm$0.2)$\times$\doce. This value is in
good agreement with the column density of 1.7$\times$\doce\, derived by \citet{McGuire2017}.

We note that HC$_7$O also has a $^2\Pi_r$ ground electronic state and  its rotational spectrum has also 
been derived from \citet{Mohamed2005}. The transitions
falling within the QUIJOTE line survey have large $J$ numbers and energies above 20\,K. This molecule
was reported as detected through stacking techniques by \citet{Cordiner2017}. These authors
assume a rotational temperature of 6\,K and obtain a column density
of (7.8$\pm$0.9)$\times$\once. Using these values, we predict an antenna temperature of 0.5 mK for
the $J$=57/2-55/2 $e$ and $f$ $\Lambda$-doubling components (31.275 GHz). 
No other lines could be detected within the
sensitivity of the QUIJOTE line survey for this rotational temperature. 
However, we have explored all the lines up to 40 GHz
and found several lines of HC$_7$O up to the $J$=69/2-67/2. They are shown in Fig. \ref{fig_hc7o}. 
Although some of the lines are blended with other lines or
with negative features resulting from the folding of the frequency switching data, the 
detection of HC$_7$O is now secure and based on the detection of several individual lines. 
The observed velocities
range from between 5.5 and 6\,\kms, which is probably due to the limited signal-to-noise ratio for
the weak emission of this species,
and to the large extrapolation in quantum numbers compared with the laboratory
data ($\nu_{max}$=19.2 GHz, $J_{max}$=35/2).

A rotational diagram indicates that the rotational temperature of HC$_7$O
could be close to 10\,K, rather than the 6\,K previously assumed \citep{Cordiner2017}. 
Using a model fitting to the observed line profiles \citep{Cernicharo2021d}
with a rotational temperature of 10\,K, we derived a column density of (6.5$\pm$0.5)$\times$\once,
which is in reasonable agreement with the value of \citet{Cordiner2017}. 
Although our data cannot support a rotational temperature as low as the value adopted by \citet{Cordiner2017}, 
the effect on the derived column density is moderate. This effect has been discussed
in detail in the analysis of the column density of HCS$^+$ and 
HC$_3$S$^+$ \citep{Cernicharo2021f}. The HC$_5$O/HC$_7$O abundance ratio is 2.2$\pm$0.3. 

We searched for HC$_4$O and HC$_6$O without success. The derived upper limits are
given in Table \ref{col_densities}.

\begin{figure}[]
\centering
\includegraphics[scale=0.74,angle=0]{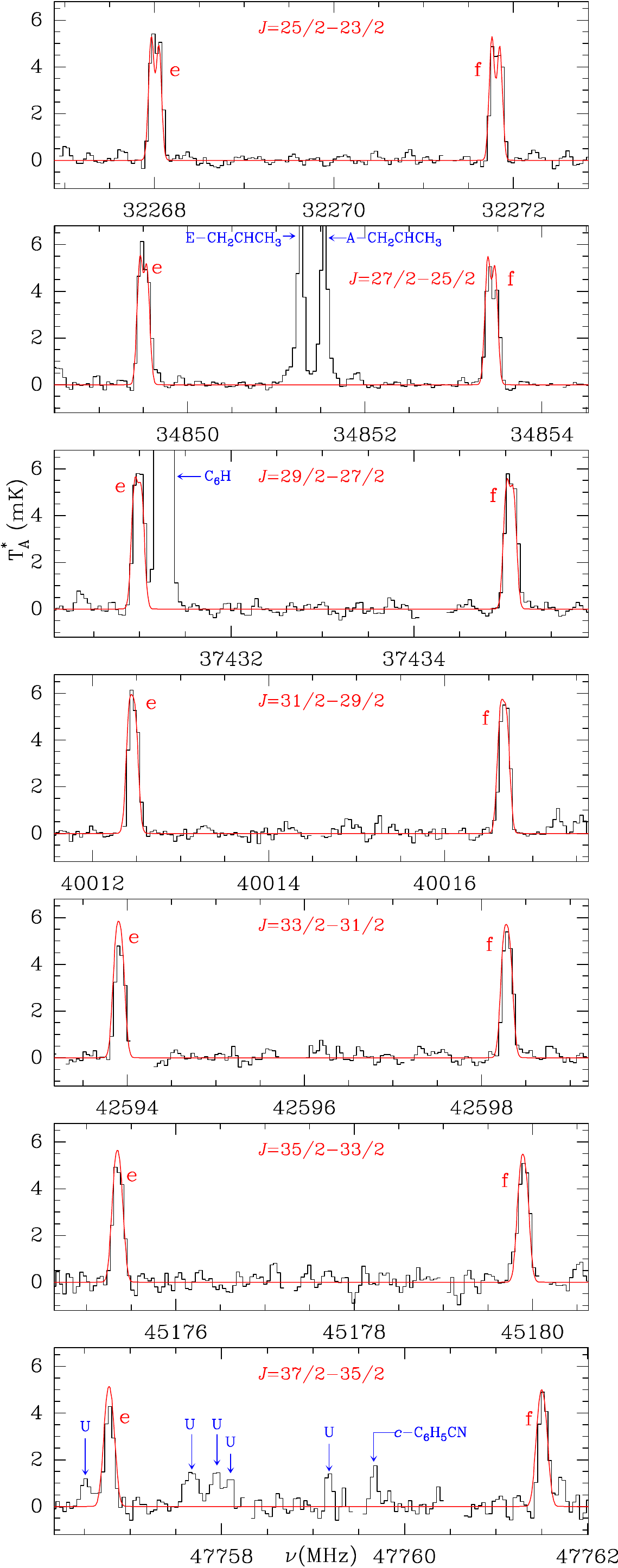}
\caption{Observed lines of HC$_5$O in TMC-1.
The abscissa corresponds to the rest frequency of the lines. The
v$_{LSR}$, linewidth, and integrated intensity of each line are given in Table \ref{obs_line_parameters}.
The ordinate is the antenna temperature, corrected for atmospheric and telescope losses, in 
milli Kelvin.
The quantum numbers for each transition are indicated. 
The red line shows the computed synthetic spectrum for this species (see Appendix \ref{Ap_HC5O_HC7O}). 
Blank channels correspond to negative features produced in the folding of the frequency
switching data. 
}
\label{fig_hc5o}
\end{figure}

\begin{figure}[]
\centering
\includegraphics[scale=0.74,angle=0]{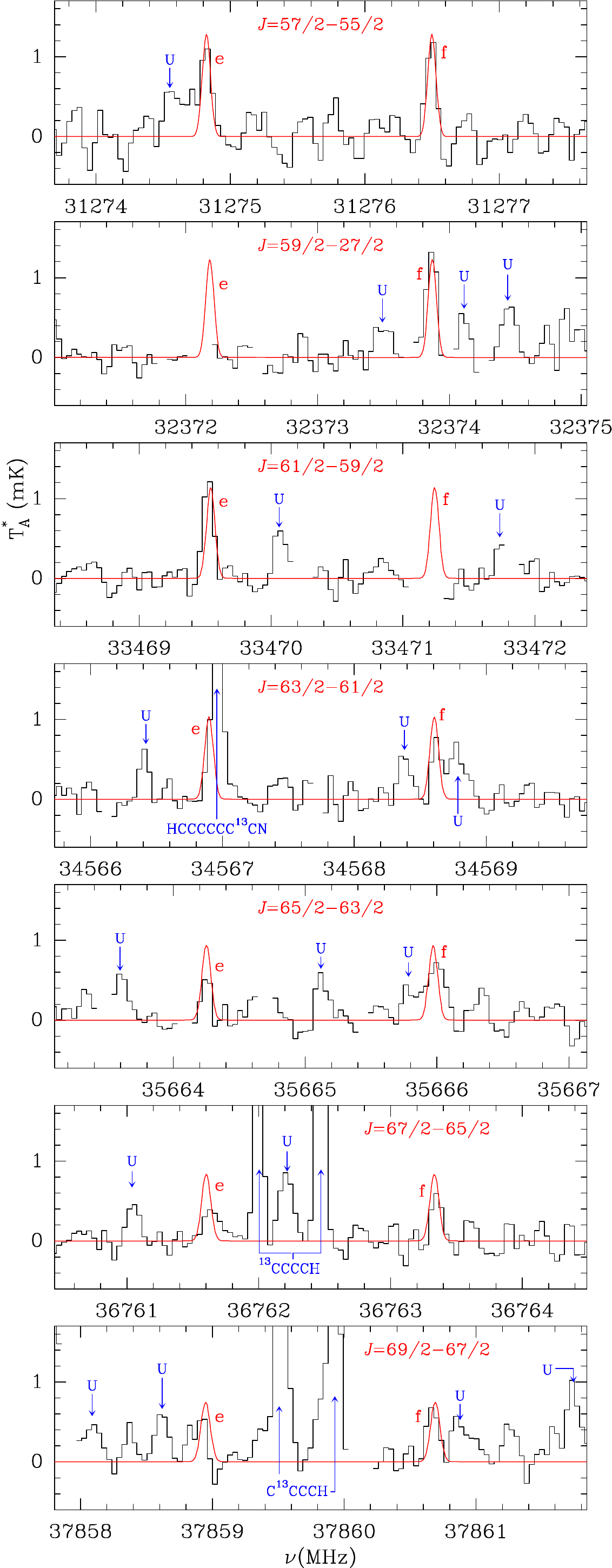}
\caption{Observed lines of HC$_7$O in TMC-1.
The abscissa corresponds to the rest frequency of the lines. The
v$_{LSR}$, linewidth, and integrated intensity of each line are given in Table \ref{obs_line_parameters}.
The ordinate is the antenna temperature, corrected for atmospheric and telescope losses, in 
milli Kelvin.
The quantum numbers for each transition are indicated. 
The red line shows the computed synthetic spectrum for this species (see Appendix \ref{Ap_HC5O_HC7O}).
Blank channels correspond to negative features produced in the folding of the frequency
switching data.  
}
\label{fig_hc7o}
\end{figure}

\begin{figure}[]
\centering
\includegraphics[scale=0.65,angle=0]{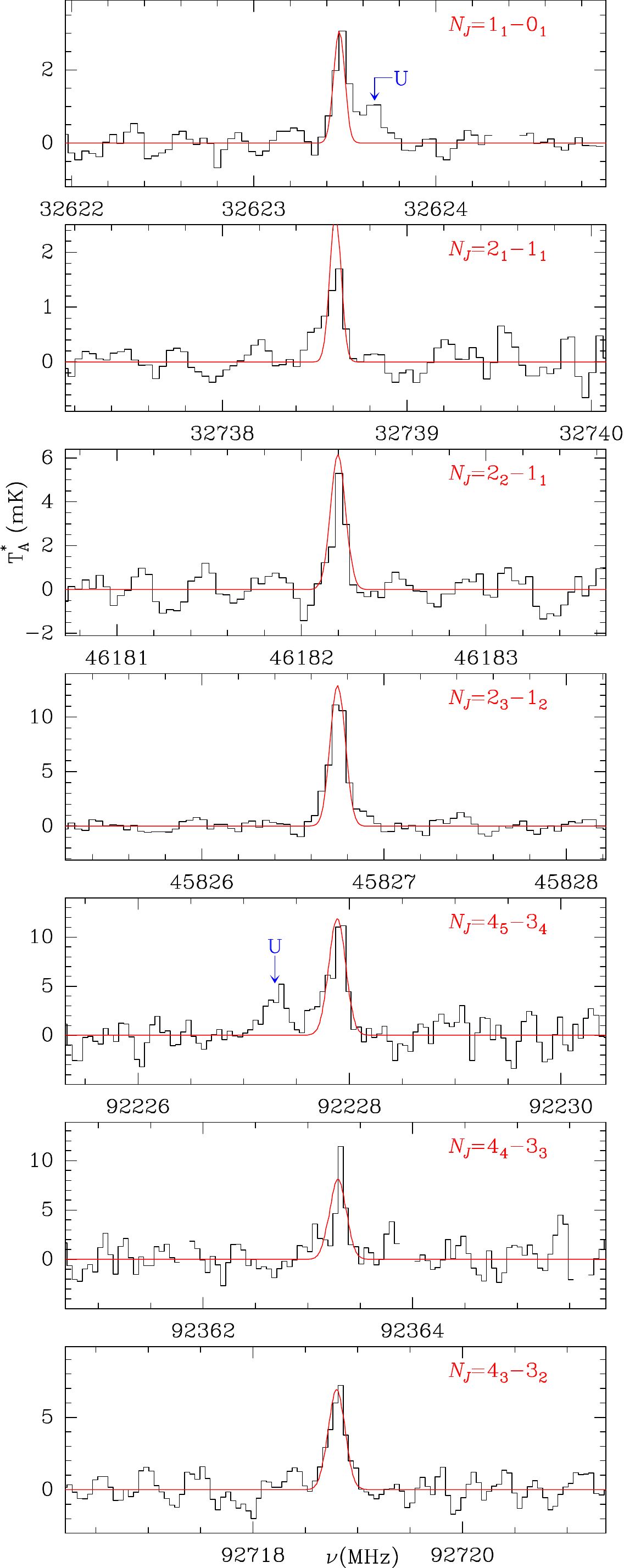}
\caption{Observed transitions of CCO in our line survey towards TMC-1 .
The abscissa corresponds to the rest frequency of the lines. The
v$_{LSR}$, linewidth, and integrated intensity of each line are given in Table \ref{obs_line_parameters}.
The ordinate is the antenna temperature, corrected for atmospheric and telescope losses, in 
milli Kelvin.
The quantum numbers for each transition are indicated
in the upper right corner of the corresponding panel.
The red line shows the computed synthetic spectrum for this species (see Appendix \ref{Ap_CCO}). 
Blank channels correspond to frequencies affected by negative features
produced in the folding of the frequency switching data.
}
\label{fig_cco}
\end{figure}

\section{C$_n$O species}
\label{Ap_CnO}
\subsection{Dicarbon Monoxide, CCO}
\label{Ap_CCO}
This molecule has a $^3\Sigma^-$ ground
electronic state and its microwave and millimetre wave spectra are well-known \citep{Yamada1985,Ohshima1995}.
We note that CCO was detected  in TMC-1 by \citet{Ohishi1991}. The laboratory data cover frequencies up to 184.8 GHz
and $J_{max}$=9, which guarantees excellent frequency predictions in the frequency
domain of our line survey. A dipole moment of 1.305\,D was derived
through ab initio calculations by \citet{Thomson1973}. The lines observed with the Yebes 40m and IRAM
30m radio
telescopes are shown in Fig. \ref{fig_cco}, and the derived line parameters for all lines are 
given in Table \ref{obs_line_parameters}. A rotational diagram assuming a source of uniform
brightness temperature and 80$''$ in diameter \citep{Fosse2001} provides a rotational temperature
of 7.0$\pm$1.5\,K and a column density of (7.5$\pm$0.3)$\times$\once. This value is in good
agreement with that obtained by \citet{Ohishi1991}, and it is slightly smaller than the one derived
by \citet{Cernicharo2020a} who assumed an uniform rotational temperature of 10\,K and did not
consider the 3mm lines. The effect of the 
adopted rotational temperature on the column density is much more critical for this $^3\Sigma^-$ molecule
than for a linear species of a similar rotational constant. This is due to the large energy difference
between the levels $J=N-1, N$, and $N+1$, and the possible propensity rules for collisions between
these levels.
The comparison of the observed intensities at 3 and 7mm suggest that at 
3mm ($N=4$) the excitation temperature is 
below 6\,K, while at low frequency it could be close to the kinetic temperature of 10\,K. 

In order
to evaluate the possible excitation effects, we used the MADEX code to perform a large 
velocity gradient calculation (LVG; MADEX follows the formalism described by \citet{Goldreich1974}). 
For this purpose, we adopted the collisional rates of
OCS \citep{Green1978} and the infinite-order-sudden approximation for a $^3\Sigma$ 
molecule \citep{Alexander1983,Corey1983,Fuente1990}. For a volume density of 4$\times$10$^4$ 
cm$^{-3}$, the excitation temperatures obtained using
this approach are $\sim$10\,K for the lines observed at low frequency (QUIJOTE data). 
However, the excitation temperature decreases very
fast to 6.5\,K for $N$=4 and to $\sim$5.5\,K for $N$=5. 
We have explored the effect of n(H$_2$) on the excitation temperatures of CCO.
For n(H$_2$)$\ge$9$\times$10$^3$ cm$^{-3}$, the $N$=1 and 2 lines have an
excitation temperature close to T$_K$ . However, the lines associated with 
the levels with $N$=4 show a sharp dependence of T$_{exc}$ with density. The best
fit to the observations is obtained for n(H$_2$)=10$^4$ cm$^{-3}$ and N(CCO)=7.5$\times$10$^{11}$
cm$^{-2}$, that is, the value derived from the rotational diagram. For this density, the excitation
temperature of the $N$=5 lines decreases to 4-5\,K.
 
The synthetic spectrum computed with the parameters resulting from the best fit
is shown in Fig. \ref{fig_cco}. 
The match between the observations and the model is rather good despite
the uncertainties on the adopted collisional rates, which also affect the derived volume density.
Collisional rates for CCO are needed in order to derive more accurate physical conditions
from its rotational transitions.

\subsection{Oxopropadienylidene, CCCO}
\label{Ap_CCCO}
There is an extensive literature on laboratory experiments concerning the centimetre, millimetre, and
sub-millimetre spectroscopy of oxopropadienylidene \citep{Brown1983,Tang1985,Klebsch1985,Bogey1986,
Bizzocchi2008}. The data cover frequencies up to 740 GHz and $J_{max}$=77. The molecule has a
dipole moment of 2.39\,D \citep{Brown1983}. This molecule has been detected in the interstellar
and circumstellar media \citep{Matthews1984,Brown1985,Tenenbaum2006}. The protonated form of this
species, HCCCO$^+$, has been recently
detected in TMC-1 \citep{Cernicharo2020a}. A study of CCO and CCCO in low-mass star-forming
regions has been recently made by \citet{Urso2019}.

The observed lines with the Yebes and IRAM radio telescopes
are shown in Fig. \ref{fig_c3o}. The derived line parameters are given in Table \ref{obs_line_parameters}.
Assuming a source of 80$''$ in diameter with a uniform brightness temperature, we obtain a
rotational temperature of 7$\pm$1\,K and a column density of (1.2$\pm$0.2)$\times$\doce. This
value is in good agreement with that reported by \citet{Matthews1984}, \citet{Brown1985}, and \citet{Cernicharo2020a}. 
The synthetic spectrum computed for these values of the rotational temperature and the
column density is shown by the red lines in Fig. \ref{fig_c3o}. The agreement between
the model and observations is excellent. In order to check
the assumption of a uniform rotational temperature for all rotational levels of CCCO,
which could be the main source error in the determination of the column density
\citep{Cernicharo2021b}, we
performed large velocity gradient calculations adopting the collisional rates of HC$_3$N \citep{Faure2016}
for CCCO.
Adopting a density of 4$\times$10$^4$ cm$^{-3}$ \citep{Cernicharo1987,Fosse2001}, we obtain an excitation 
temperature of $\sim$10\,K for all levels below $J\le$5, and a continuous decrease in $T_{exc}$ with
increasing $J$, reaching a value of $\sim$6\,K for $J=11$. Hence, the derived rotational temperature is strongly
determined by the transitions with high-$J$ levels observed with the IRAM 30m radio telescope.

\begin{figure}[]
\centering
\includegraphics[scale=0.65,angle=0]{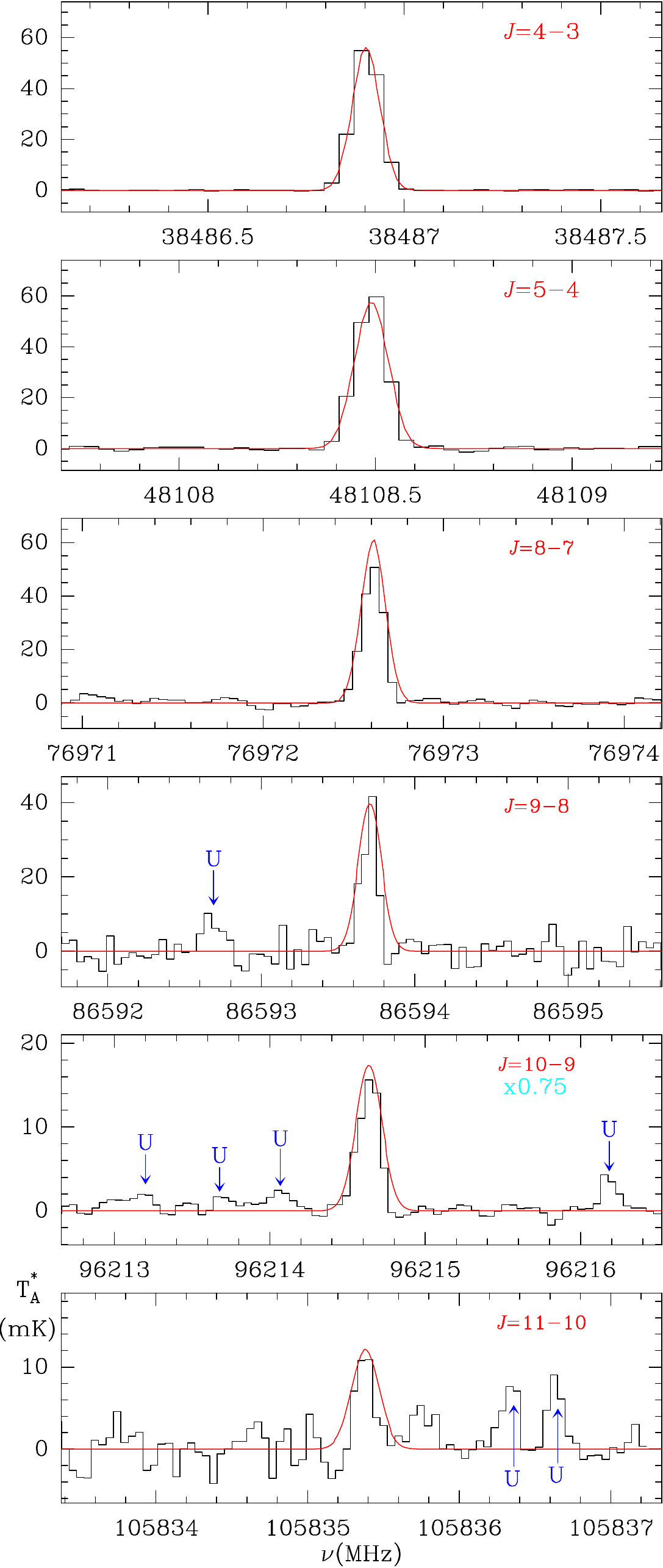}
\caption{Observed transitions of C$_3$O in TMC-1.
The abscissa corresponds to the rest frequency of the lines. Their
v$_{LSR}$, linewidth, and integrated intensity of each line are given in Table \ref{obs_line_parameters}.
The ordinate is the antenna temperature, corrected for atmospheric and telescope losses, in 
milli Kelvin.
The quantum numbers for each transition are indicated
in the upper right corner of the corresponding panel.
The red line shows the computed synthetic spectrum for this species (see Appendix \ref{Ap_CCCO}). 
}
\label{fig_c3o}
\end{figure}

\subsection{C$_4$O and C$_6$O}
\label{Ap_C4O_C6O}
We note that C$_4$O and C$_6$O were observed in the laboratory by \citet{Ohshima1995}. Both have
a $^3\Sigma^-$ ground electronic state. These molecules have been implemented in MADEX
and we searched for their strongest transitions within the QUIJOTE line survey. None 
of their lines are detected. Assuming a rotational temperature identical to
that of C$_5$O (T$_{rot}$=10\,K), we derived 3$\sigma$ upper limits to their column density
of 9.0$\times$10$^{10}$ and 1.1$\times$10$^{11}$ cm$^{-2}$,
respectively.

\end{appendix}

\end{document}